\documentstyle[11pt,psfig]{l-aa}
\begin{document}

\newcommand{\muz}{\mu_0}
\newcommand{\epz}{\varepsilon_0}
\newcommand{\va}{Alfv\'en speed\/ }

\newcommand{\smallfrac}[2]{{\small \frac{#1}{#2}}}

\author{N. Schutgens}
\thesaurus{09(02.13.2; 06.03.2; 06.13.1; 06.15.1; 06.16.2; 06.16.3)}

\title{Numerical simulation of prominence oscillations}

\author{N.A.J. Schutgens\inst{1}, G. T\'{o}th\inst{2}}

\institute{Royal Netherlands Meteorological Institute ({\scshape knmi}), De Bilt, The Netherlands
\and Department of Atomic Physics, E\"{o}tv\"{o}s University, Hungary}

\offprints{Nick Schutgens, {\scshape schutgen@knmi.nl}}

\date{}

\maketitle

\begin{abstract}
We present numerical simulations, obtained with the Versatile Advection Code, of
the oscillations of an inverse polarity prominence.
The internal prominence equilibrium, the surrounding corona and the inert photosphere 
are well represented.
Gravity and thermodynamics are not taken into account, but it is argued that
these are not crucial.
The oscillations can be understood in terms of a solid body moving through
a plasma. The mass of this solid body is determined by the magnetic field topology, not by the
prominence mass proper. The model also allows us to study the effect of the ambient coronal plasma on the motion of the
prominence body. Horizontal oscillations are damped through
the emission of slow waves while vertical oscillations are damped through the emission of
fast waves.
\end{abstract}

\section{Introduction}
\label{se:intro}

Solar prominence oscillations have been the subject of both observational and theoretical papers for the
past 35 years. One of the first studies (Ramsey \& Smith 1966) 
concerned observations of global oscillations of disk filaments with periods of $6^m$ to $40^m$, that were interpreted
by Hyder (1966) as predominantly vertical motions and by Kleczek \& Kuperus (1969) as predominantly
horizontal motions.

Tsubaki (1988) published a review on oscillation studies of limb prominences. Most of these oscillations
pertain to Doppler shifts in the spectra of part of a prominence. The associated mass flows are essentially parallel to the photosphere,
both
longitudinal and transverse to the prominence main axis. The observed periods range from $160^s$ to
$82^m$, with velocity amplitudes in the range of 0.2--3 km/s. 

Zhang et al.\  (1991), Zhang \& Engvold (1991) and Thompson \& Schmieder (1991) studied disk
filaments and found locally periods of $2.5^m-22^m$, with velocity amplitudes of $0.5-1.25$ km/s. 
As in the observations by Ramsey \& Smith these oscillations may have both horizontal and vertical components. 

Since Tsubaki's review more limb studies have been performed by Mashnich \& Bashkirtsev (1990),
Suematsu et al.\  (1990),
Bashkirtsev \& Mashnich (1993), Mashnich et al.\  (1993),
Balthasar et al.\  (1993),
Balthasar \& Wiehr (1994), Park et al.\  (1995),
S\"utterlin et al.\  (1997) and Molowny-Horas et al.\  (1997). These observations all confirm and
extend previous results. There is now enough evidence to suggest  the following
tentative classification for prominence oscillations (see also Bashkirtsev \& Mashnich 1993).
\begin{itemize}
\item very short periods: $P \approx 30^s$ (Balthasar et al.\  1993), perhaps due to fast 
waves propagating along flux tubes (Roberts et al.\  1984).
\item short periods: $P \approx 3^m-10^m$, at least some of which are related to
photospheric or chromospheric forcing with periods of $3^m$ and $5^m$ (Balthasar et al.\  1986, Zhang et al.\ 
1991).
\item intermediate periods: $P \approx 10^m-40^m$, which are probably genuine eigenmodes
of the prominence (Ramsey \& Smith 1966, Balthasar et
al. 1988, Bashkirtsev \& Mashnich 1993). Many prominences show nearly the same oscillation period
each time they are perturbed. 
\item long periods: ($P \approx 40^m-114^m$), which may be related to chromospheric
forcing (Balthasar et al.\  1988).
\end{itemize}
We point out that the longest data set is about 7 hours long and the best temporary
resolution is a couple of seconds.

In particular the observed $10^m-40^m$ eigenmodes are interesting as they are damped and thus loose
energy, perhaps due to some
interaction with the ambient corona (Ramsey \& Smith 1966, Kleczek \& Kuperus 1969). 
Typically, the quality factor $Q=\pi T_{\rm damp}/P< 6$, with $T_{\rm damp}$ the
damping time of the oscillation. This implies that three to four oscillations
can be observed
after the impulsive perturbation of the prominence.
Understanding the damping mechanisms will give more insight in prominence dynamics and
may yield an extra diagnostic tool for prominence
and ambient coronal plasma parameters.

Prominence oscillations have been theoretically modelled by many authors.  Some model the
prominence as a harmonic oscillator with an ad-hoc
damping term: Hyder (1966) used viscous effects, Kleczek \& Kuperus (1969) used
emission of sound waves, van
den Oord \& Kuperus (1992), Schutgens (1997a) and van den Oord et al.\  (1998) used
emission of Alfv\'en waves.
Other authors construct a simple MHD equilibrium and study oscillations thereof using the 
linear adiabatic MHD equations
(Oliver et al.\  1992, 1993, Oliver \& Ballester 1995, 1996, Joarder \& Roberts 1992ab, 1993,
Joarder et al.\  1997). Generally the latter approach yields marginally stable oscillations (no
damping), although  Joarder \& Roberts (1992a) and Joarder et al.\  (1997) claim that leaky waves are possible solutions
to their equations. 
Apparently, little effort has gone into studying the damping mechanisms themselves.

Schutgens (1997ab) and van den Oord et al.\  (1998) recently studied  the
global equilibrium of  prominences treating the evolution of the magnetic field in a
self-consistent way. The equation of motion for the prominence (approximated as a line current) was
solved simultaneously with the Maxwell equations for the electro-magnetic fields. Their results
show that the Alfv\'en travel time between prominence and photosphere $\tau$ is an important time
scale of the system. In particular, van den Oord et al.\  found that it is only possible to obtain {\em
stable} prominence equilibria by taking damping mechanisms into account. Hence, damping is not
just necessary to describe prominence oscillations quantitatively correct, but it is an essential
ingredient of a prominence equilibrium.  

In this paper, we study prominence oscillations, and in particular the effect of the
ambient coronal plasma on the prominence motion. We use the Versatile Advection
Code (VAC) to solve numerically the isothermal MHD equations in two
dimensions. VAC has been developed by G. T\'{o}th (1996, 1997) and is capable of solving a
variety of hydrodynamical and magneto-hydrodynamical problems in one, two and
three dimensions using a host of
numerical methods. 
In Sect.\ \ref{se:wire} a simple analytical model for prominence equilibrium and
dynamics is discussed that will be used for comparison with our numerical results. In section\
\ref{se:nummethod} we describe the 
isothermal equations that were solved numerically, the methods used,
the grid structure and the initial conditions. In Sect.\ \ref{se:2dresults} the 
simulations are described. Using the analytical model mentioned before, these simulations are
interpreted in terms of the physical processes involved. A summary and the
conclusions can be found
in Sect.\ \ref{se:summary}. All dimensional variables in this paper are in rational MKSA units,
unless stated differently.

\section{The line current approximation}
\label{se:wire}
We
briefly recapitulate the line current  approximation for prominence equilibrium and dynamics. 
This approximation will serve as a guideline for discussing our numerical results.
The prominence is 
approximated by a straight, infinitely thin and long, line current $I_0>0$ at a height $y_0$
parallel to the photosphere. Along the prominence we assume invariance. The effect of the massive photosphere on the prominence magnetic
field is modelled through a mirror current $-I_0$ at depth $-y_0$ below the surface of the Sun
(Kuperus \& Raadu 1974, van Tend \& Kuperus 1978, Kaastra 1985, Schutgens 1997a).
The momentum equations governing the global prominence dynamics are (ignoring gravity)
\begin{eqnarray}
\sigma \ddot{x} & = -&I_0 \left[ B_{\rm mir}^y(x,y) + B_{\rm cor}^y(x,y) \right] -\nu_x 
\dot{x}, \nonumber \\
 \sigma \ddot{y} & = &I_0 \left[ B_{\rm mir}^x(x,y) + B_{\rm cor}^x(x,y) \right] -\nu_y 
\dot{y}.
\label{eq:eqofmot}
\end{eqnarray}
where $\sigma$ is the longitudinal mass density of the oscillating structure, $\vec{B}_{\rm
cor}$ is the coronal arcade field in which the prominence is located and $\vec{B}_{\rm
mir}$ is the field due to the mirror current. 

The interaction
between the {\em moving} filament and the ambient coronal plasma gives rise to
viscous effects (Hyder 1966) and emission of magneto-acoustic waves (Kleczek \&
Kuperus 1967, van den Oord \& Kuperus 1992) that act as damping
mechanisms. These are heuristically modelled through $\nu_x$ and $\nu_y$,
damping constants that are in reality determined by the flow field
around the prominence. Since viscosity of the coronal plasma is negligible, 
we concentrate on the emission of magneto-acoustic waves. An approximation for the
damping constants can be found by considering linear motion with constant velocity $v$ of a solid body through a homogeneous
plasma. 
If the cross section of the body perpendicular to its motion is $A$, the body
transfers $2 \rho_{\rm cor} c A v $ momentum per unit time onto the plasma
(Landau \& Lifschitz
1989, p. 256).
Here $c$ is the characteristic wave speed of the plasma with density $\rho_{\rm
cor}$. A factor $2$ is added since both the front-
and backside of the object transfer momentum. Hence, the damping constants have the form
\begin{equation}
\label{eq:damping}
\nu=2 \rho_{\rm cor} c A.
\end{equation}
\noindent
The coronal arcade is generated by a magnetic line dipole $M_{\rm d}>0$, a depth
$H_{\rm d}$ below the photosphere
\begin{eqnarray}
B_{\rm cor}^x(x,y) & = & \frac{\muz M_{\rm d}}{\pi} 
\frac{x^2-(y+H_{\rm d})^2}{\left( x^2 + (y+H_{\rm d})^2 \right)^2}, \nonumber \\
B_{\rm cor}^y(x,y) & = & \frac{2 \muz M_{\rm d}}{\pi}
\frac{x (y+H_{\rm d})}{\left( x^2 + (y+H_{\rm d})^2 \right)^2}.
\label{eq:arcadefield}
\end{eqnarray}
\noindent
The mirror current's field at the location of the filament, in the
quasi-stationary approach, 
is given by
\begin{eqnarray}
B_{\rm mir}^x(x,y) & = & \frac{\muz}{4 \pi} \frac{I_0}{y}, \nonumber \\
B_{\rm mir}^y(x,y) & = & 0.  
\label{eq:mirrorfield}
\end{eqnarray}
The fact that $B_{\rm mir}^y=0$ is a direct consequence of the mirror current
mirroring the motion of the filament (to conserve the photospheric flux) and the
quasi-stationary field assumption. When this assumption is dropped and
the magnetic fields evolve dynamically according to Maxwell's laws, the
expressions for $\vec{B}_{\rm mir}$ become far more complicated and in
particular $B_{\rm mir}^y \neq 0$ (Schutgens 1997a, van den Oord et al.\  1998).

\begin{figure}[htb]
\centerline{\psfig{figure=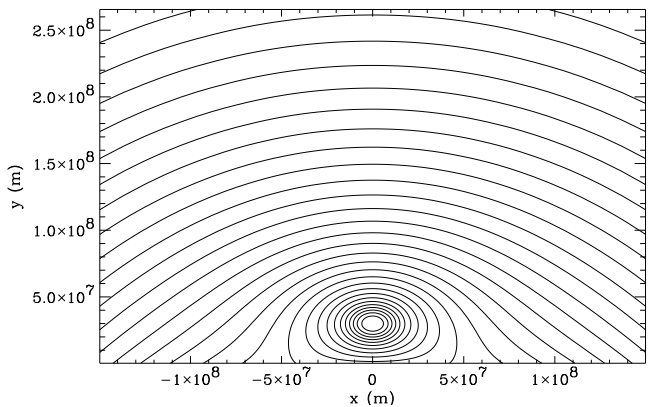,width=8.6cm,bbllx=50pt,bblly=373pt,bburx=550pt,bbury=700pt,clip=t}}
\center{\caption{\protect\footnotesize The magnetic field topology after relaxation. A
strong, localized current runs near $(x,y)=(0,3\times\, 10^7)$ m, at the location of the
prominence.
\label{fi:topol}}}
\end{figure}

Assuming quasi-stationary field evolution ($v_{\rm A} \rightarrow \infty$),
prominences are in stable equilibrium provided they are
on the symmetry axis of the arcade ($y=0$)
and at a height $y_0 < H_{\rm d}$. Furthermore, the current should attain the value
\begin{equation}
I_0= \frac{4 y_0 M_{\rm d}}{(y_0+H_{\rm d})^2}.
\label{eq:equilibrium}
\end{equation}
Note that this prominence equilibrium has an inverse polarity topology (see also Fig.\
\ref{fi:topol}). In fact, it corresponds to a Kuperus-Raadu prominence (see van Tend
\& Kuperus 1978).
\begin{figure}[htb]
\centerline{\psfig{figure=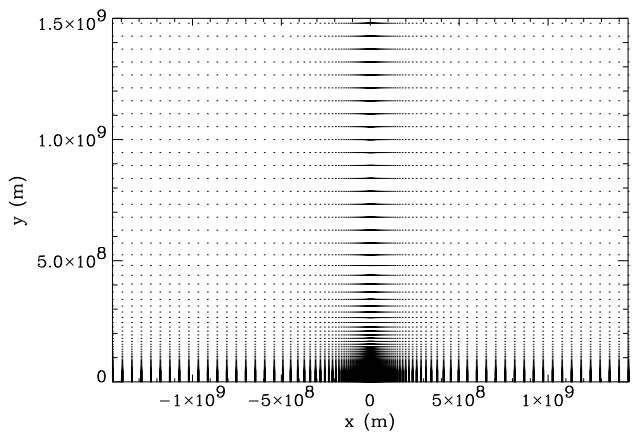,width=8.6cm}}
\center{\caption{\protect\footnotesize Location of cell centers of the
grid. The photosphere is at $y=0$,
and the prominence at $(x,y)=(0,3\times 10^7)$ m. Its radius is $1.5\times\,
10^7$ m. The complete grid is $3\times\,10^9$ m by $1.5\times\,10^9$ m.
\label{fi:grid}}}
\end{figure}

If one linearizes around this equilibrium, the equations of vertical and
horizontal motion decouple
(due to the symmetry of the coronal field) and both have the form of the familiar damped harmonic
oscillator.
The solutions are oscillations, characterized by frequencies $\omega$ and 
damping rates $\delta$:
\begin{eqnarray}
\omega_x  & = & \left( \Omega^2_x - \frac{1}{4} \left( \frac{\nu_x}{\sigma} \right)^2 \right)^{\frac{1}{2}}
\quad \Omega^2_x=\frac{8 \muz M_{\rm d}^2}{\pi \sigma} \frac{y_0}{(y_0+H_{\rm
d})^5}, \label{eq:horfreq} \\
\delta_x & = & -\frac{\nu_x}{2 \sigma} \label{eq:hordamp}, \\
\omega_y  & = & \left( \Omega^2_y - \frac{1}{4} \left( \frac{\nu_y}{\sigma} \right)^2 \right)^{\frac{1}{2}} 
\quad \Omega^2_y=\frac{4 \muz M_{\rm d}^2}{\pi \sigma} \frac{H_{\rm d}-y_0}{(y_0+H_{\rm
d})^5}, \label{vertfreq} \\
\delta_y & = & -\frac{\nu_y}{2 \sigma} \label{eq:vertdamp}.
\end{eqnarray}
where $\Omega_x$ and $\Omega_y$ are the quasi-stationary frequencies of the system {\em without} damping.

If one drops the quasi-stationary assumption, the same equilibrium is found, but its stability
then also depends on the value of the coronal \va. In general, the solutions to
the equations,
which are again decoupled, are damped or growing (!) harmonic oscillations
(Schutgens 1997ab, van den Oord et al.\  1998). 

\section{Numerical methods}
\label{se:nummethod}
We solve the time-dependent ideal isothermal MHD equations in two dimensions using the Versatile
Advection Code (VAC) developed by one of us (G. T\'{o}th). Descriptions of this numerical code can be
found in T\'{o}th (1996, 1997).
The
photosphere coincides with the $y=0$ plane. In conservative form the equations for density
$\rho$, mass flux $\rho \vec{v}$ and magnetic field $\vec{B}$
are ($p$ is the thermal pressure)
\begin{eqnarray}
\partial_t \rho  & + \nabla \cdot \left(\vec{v} \rho \right) & =0, \nonumber \\
\partial_t (\rho \vec{v}) & + \nabla \cdot \left( \vec{v} \rho \vec{v} - \vec{B} \vec{B} \right)
 + \nabla
(p + B^2/2) & = 0, \nonumber \\
\partial_t \vec{B} & + \nabla \cdot \left( \vec{v} \vec{B} - \vec{B} \vec{v}
\right) & =0.
\label{eq:isomhd}
\end{eqnarray}
Note that we ignore gravity (see Sect.\ \ref{se:summary} for a discussion).
These equations must be solved together with an equation of state $p=c^2_{\rm s} \rho$ and the condition
that $\nabla \cdot \vec{B}=0$. Here $c_{\rm s}$ is the sound speed, a free parameter of the
system, which is constant throughout the numerical domain.
The magnetic field unit is chosen such that the current density satisfies $\vec{J}=\nabla \times
\vec{B}$ (i.e. $\mu_0=1$), all other units are rational MKSA. 
\begin{figure}[htb]
\centerline{\psfig{figure=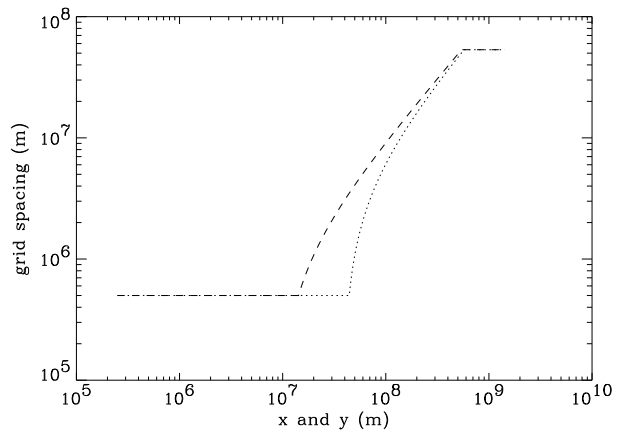,width=8.6cm}}
\center{\caption{\protect\footnotesize Grid spacing as a function of position in
the grid.
The dashed line represents the resolution 
in the $x$-direction, the dotted line in the $y$-direction. The photosphere is at $y=0$,
and the prominence at $(x,y)=(0,3\times 10^7)$ m. 
\label{fi:gridspacing}}}
\end{figure}

The equations are discretized on the same grid and solved using a FCT (Flux
Corrected Transport)
scheme. Since there are no discontinuities in the solution, FCT performs well.
Similar results can be obtained using a TVD (Total Variation Diminishing) Lax-Friedrichs scheme but, for
the problem at hand, FCT is more efficient.  
For a comparison of different methods see T\'{o}th \& Odstr\v cil (1996).
A projection scheme (Brackbill \& Barnes 1980) is used to keep the magnetic divergence
small. Typically $|\nabla \cdot \vec{B} | < 10^{-3} B/L$, where $B$ and $L$ are
characteristic strength and length scale of the magnetic field. 
\begin{figure*}[htb]
\centerline{\psfig{figure=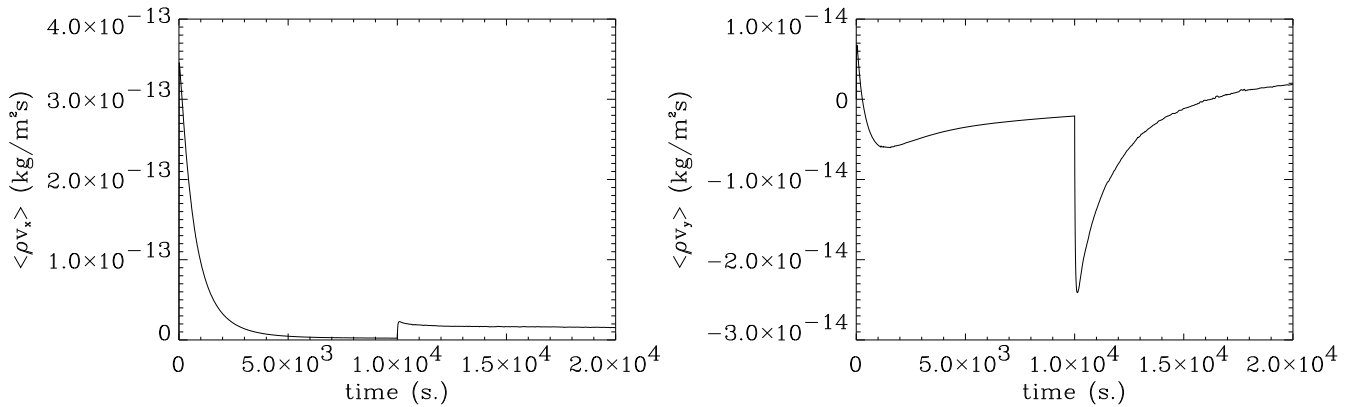,width=18cm}}
\center{\caption{\protect\footnotesize The volume averaged horizontal and
vertical momentum: $\rho v_{x,y} = \int_V \rho v_{x,y} \, dV / V$, $V=3\times\, 10^9$ m $\times\,
1.5\times\, 10^9$m.
Note the temporary increase at $t=10\,
000$ as the artificial damping factor $\alpha$ is changed. 
\label{fi:relax}}}
\end{figure*}

We use a Cartesian grid of $190 \times 155$ cells ($3\times\,10^9$ m by $1.5\times\,10^9$ m)
that is strongly distorted, with the highest resolution at the location of the
prominence and a
much smaller resolution near the coronal edges of the computational
domain (see Fig.\ \ref{fi:grid}). In a region surrounding the actual prominence
the grid spacing is constant.
Typically, the grid spacing increases with 10\% from cell to cell outside this region. As a
consequence the difference in resolution at the prominence and at the far
coronal edges can amount to a factor 100 (see Fig.\ \ref{fi:gridspacing}).
Near the boundaries the grid spacing is again kept constant.

The boundary conditions are implemented using two layers of ghost cells around the 
physical part of the grid. The actual boundary is 
located between the inner ghost cells and the outermost cells of the physical grid. 
The solution on the physical part of the grid can be advanced using fluxes
calculated from the ghost and physical cells around the boundary.  
The prescription for the
ghost cells depends on the specific boundary condition and may be constant or
depend on the solution in the adjacent physical cells. 

The photosphere is much denser than the corona and is therefore  
strongly reflecting: there should be no mass flux or
energy flux across it. 
We therefore choose the plasma density symmetric around the photospheric boundary
($\rho^{\rm ghost} =\rho^{\rm physical}$) and vertical momentum anti-symmetric
($\rho v_y^{\rm ghost} =-\rho v_y^{\rm physical}$).
The other momentum component $\rho v_x$ is chosen
anti-symmetric as well, since we assume no flows along the photosphere (`no
slip'
condition).
The magnetic field is tied to the dense photosphere and the photospheric flux
($B_y$) is
conserved. Magnetic waves should be reflected.
The initial field solution 
is stored in computer memory and subtracted from the advanced
solution. The `linearized' field solution thus obtained is chosen to be 
symmetric or anti-symmetric in the photospheric boundary: 
$B_x^{\rm ghost} = B_x^{{\rm ghost},0}+(B_x^{\rm physical}-B_x^{{\rm physical},0})$  and
$B_y^{\rm ghost} = B_y^{{\rm ghost},0}-(B_y^{\rm physical}-B_y^{{\rm physical},0})$.
In this way, one obtains a good reflection of waves for small perturbations. 

The other (coronal) boundaries do not coincide with a physical boundary and
should be as open as possible.
Since any choice of boundary condition will always generate some
reflection, which we want to avoid, we decided to place these
boundaries at large distances from the prominence so that twice the wave crossing time
(prominence--coronal boundary) is longer than the simulation time.
Also, the coarsening of the grid damps the outward moving waves and the
reflection is minimized.
At the location of the coronal boundaries we prescribe fixed  $\vec{B}^{\rm ghost}$, but 
copy $\rho$ and $\rho \vec{v}$ from  the physical part of the grid to the ghost cells.

The initial configuration is a superposition of three different
analytical MHD equilibria. The global coronal structure is
a potential arcade, given by Eq.\ (\ref{eq:arcadefield}).
Since we ignore gravity, the plasma density in this arcade is constant.
To this equilibrium we add the fields and plasma densities of two current
carrying flux tubes, one  above, the other below the photosphere. 
Both flux tubes are located on the polarity inversion line of the arcade. The
flux tube at a height $y_0$ {\em above\/} the photosphere represents the prominence.
Its total axial current is $I_0$. The flux tube $-y_0$ {\em below} the photosphere represents
the mirror current and has a total current $-I_0$. 

The flux tube equilibrium is derived starting from a current profile (see also Forbes 1990)
\begin{equation}
j_z(r) = \left\{
\begin{array}{lc}
  j_0  & \quad\quad\quad r \le r_0-\Delta r_0,  \\
   j_0  \sin^2 \left( \frac{\pi}{2} \frac{r-r_0}{\Delta r_0} \right) &
	 \quad\quad\quad  r_0-\Delta r_0 < r \le r_0,  \\
 0 & \quad\quad\quad r > r_0. 
\end{array}
\right.
\end{equation}

\noindent
The radius of the tube $r_0$ is larger than $\Delta r_0$, the size of the region
in which the current density drops off to zero.
The total axial current in the flux tube is
\begin{equation}
I_0 =2 \pi \int_0^{r_0} r' j_z(r') \, dr'
\end{equation}
\noindent
and the  azimuthal field component follows from Stokes' theorem
\begin{equation}
B_\phi(r) = \left\{ \begin{array}{lc} 
\frac{\mu_0}{r} \int_0^{r} r' j(r') \, dr' & \quad\quad\quad\quad\quad\quad r \le
r_0, 
\\
 \frac{\mu_0 I_0}{2 \pi r}  & \quad\quad\quad\quad\quad\quad r > r_0.  
\end{array}
\right.
\end{equation}

\noindent
The current is confined to $r< r_0$. Beyond this range, the field is potential.
The pinching field $B_\phi$ has to be balanced by gas pressure
\begin{equation}
p(r)=-\int_{r_0}^r B_\phi(r') j(r) \, dr,
\end{equation}
where it is assumed that the flux tube proper carries no mass outside $r> r_0$.
The longitudinal mass density of the flux tube equals
\begin{equation}
\sigma_0= 2 \pi \int_0^{r_0} r' p(r')/c_{\rm s}^2\, dr',
\end{equation}
\noindent
while the total longitudinal mass density of the prominence is (due to the superposition of equilibria)
\begin{equation}
\sigma=\sigma_0+\pi
\rho_{\rm cor} r_0^2.
\end{equation}

\noindent
In a global sense,
equilibrium is obtained when Eq.\ (\ref{eq:equilibrium}) is satisfied.
However, 
especially near the outer edge of the prominence no force balance exists.
It is necessary to let the initial configuration relax to a numerical equilibrium,
before studying oscillations. As it turns out, force imbalance is mostly due to
the gradients in density and field {\em within\/} the flux tube (i.e.
numerical discretization errors) and a stable equilibrium is readily obtained.
To prevent the flows at the outer edge of the prominence from becoming too large
during the relaxation phase, and cause an instability, an artificial damping
term $\partial_t \rho \vec{v}=-\alpha
\rho \vec{v}$ is used. This term is switched off during subsequent oscillation studies.
\begin{figure}[htb]
\centerline{\psfig{figure=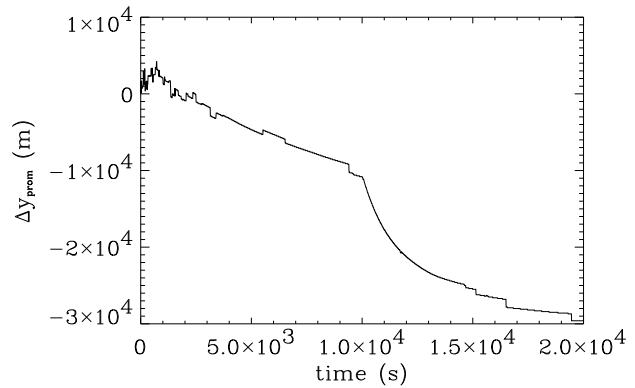,width=8.6cm}}
\center{\caption{\protect\footnotesize The change in height of the prominence during relaxation.
Initially the prominence was at $y_0=3\times\, 10^7$ m. During relaxation the
prominence moves downward a mere $0.1$ \% of its initial height.
\label{fi:relaxy0}}}
\end{figure}

Once a numerical equilibrium has been found, we perturb it and study the
resulting oscillations. In all cases, the perturbation is caused
by instantaneously adding momentum to the prominence. 

To check the validity of our results, we performed oscillation simulations for the same 
physical parameters, but different
numerical parameters. In particular we changed the grid resolution ($\Delta 
x=5\times 10^5$ or  $10^6$ m inside the prominence), the amount
of stretching of the grid (0\%, 5\% and 10\%), the duration of the relaxation ($T=9\,000$ or
$T=20\,000$ s) and the
constraint on $ | \nabla \cdot \vec{B} | $ ($<10^{-4},10^{-3}\, B/L$) and
found that the results are similar. Using $0\%$ stretching implies that the grid
is rather small in its physical size, due to computational limitations.
Reflection at the coronal boundaries will then limit the usefulness of the
simulations to the first $4000$ s for $\rho_{\rm cor}=10^{-12}$ kg/m$^3$.

\begin{figure*}[hbt]
\centerline{\psfig{figure=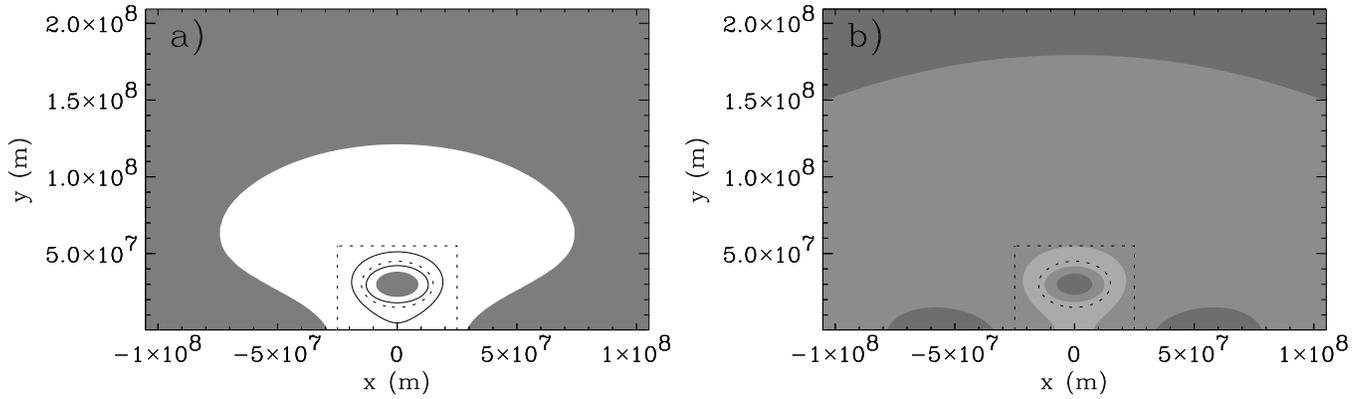,width=18cm}}
\center{\caption{\protect\footnotesize The plasma-$\beta$ (a) and \va (b) in and near the
prominence. 
In the left graph, gray denotes $\beta > 1$. The solid line is the contour line $\beta=0.1$. 
In the right graph, darkest gray denotes $v_{\rm A} < c_{\rm s}$ and lightest
gray denotes $v_{\rm A} > 500$ km/s. Maximum \va is $873$ km/s. 
The dotted circle represents the
boundary of the prominence, the dotted rectangles the coronal surface over
which the Poynting flux was integrated. The graphs are valid for $\rho_{\rm cor}=10^{-12}$ kg/m$^3$. For
other coronal densities the scalings $\beta \propto \rho_{\rm cor}$ and $v_{\rm
A} \propto 1/\sqrt{\rho_{\rm cor}}$  can be used.
\label{fi:betava}}}
\end{figure*}
In addition, the effect of the photospheric boundary condition was studied.
A perfectly reflecting photosphere does not
allow a net momentum or energy flux. Also, the magnetic flux distribution
is constant. The implementation of the photospheric boundary conditions automatically ensures
zero momentum flux and constant magnetic flux. However, since
the magnetic field is not symmetric in $y=0$, a small Poynting flux is present.
Consider a box-shaped surface around the prominence.
The coronal `walls' of this surface each have a shortest distance
to the prominence center of $2.5\times 10^7$ m (see Fig. \ref{fi:betava}) and reach down to the
photosphere, along which we close the box. The total time-averaged Poynting flux
through the photosphere is typically $10^{-3}-10^{-2}$ times smaller than the
coronal flux, for vertical oscillations. For horizontal oscillations, the
coronal flux is  small in essence (due to the anti-symmetry) and provides no
reasonable yardstick (of course one could still use the flux through a single
`wall', but this is bound to lead to similar results as for vertical
oscillations). Furthermore, we point out that the equilibrium height of the
prominence during oscillation studies does not change by
more than 0.06\%, and often even less. 
We therefore feel that the photospheric boundary is sufficiently well
represented in our numerical boundary conditions. 

\section{Oscillation studies}
\label{se:2dresults}

Before we study oscillations in prominences, we first have to compute a stable,
numerical prominence equilibrium. 
\begin{figure*}[hbt]
\psfig{figure=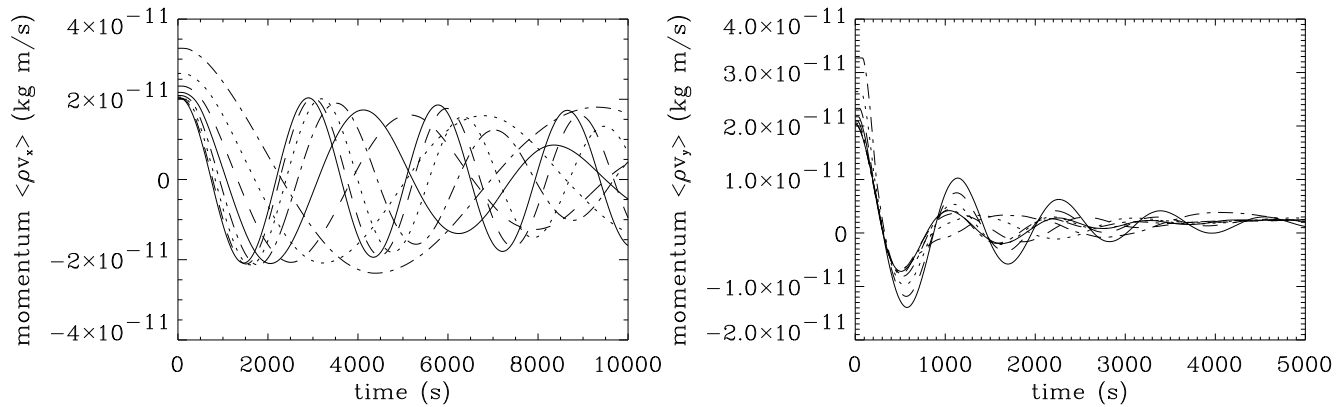,width=18cm}
\center{\caption{\protect\footnotesize The volume averaged horizontal and vertical momenta $\rho
v_{x,y}$ as a
function of time for all eight simulations. Curves with a higher initial value of
the momentum correspond to larger coronal densities. All simulations were
started with the same {\em velocity} perturbation (10 km/s). For the sake of
clarity only the first $5000$ s are shown for the vertical momentum.
\label{fi:allm}}}
\end{figure*}

\begin{table*}[hbt]
\begin{center}
\caption[]{Characteristic time scales of prominence dynamics for all eight
cases considered. The coronal density $\rho_{\rm cor}$ and (as a consequence) the prominence longitudinal mass
density $\sigma$ are the only parameters 
that differ. In particular the total current $I_0=6.5\times \, 10^{10}$ A and
the initial perturbation ($10$ km/s) are the same for all simulations.
The estimated error in the period is typically less than $3 \%$, vertical periods having larger errors than horizontal
periods. The estimated error in the damping time is typically less than $10 \%$, vertical damping times having smaller errors than horizontal
damping times. For the horizontal oscillations, we used only cases 1--6, for
vertical oscillations only cases 1--5 to compare the numerical results to the
simple line current model of Sect.~\ref{se:wire}. No reliable vertical time scales could be
obtained for the last two cases due to strong transients and noise.
\label{ta:2dresults}}
\begin{tabular}{lrrrrrrr}
\noalign{\smallskip}
\hline
\noalign{\smallskip}
    &  \multicolumn{1}{c}{coronal density} & \multicolumn{1}{c}{prominence} &
	 \multicolumn{1}{c}{effective}& \multicolumn{2}{c}{horizontal oscillations} &
	 \multicolumn{2}{c}{vertical oscillations} \\
Nr. & \multicolumn{1}{c}{$\rho_{\rm cor}$ (kg/m$^3$)} & \multicolumn{1}{c}{mass
$\sigma$ (kg/m)} & \multicolumn{1}{c}{mass
$\sigma_{\rm eff}$ (kg/m)}& \multicolumn{1}{c}{period (s)} & \multicolumn{1}{c}{damping
(s)}  & \multicolumn{1}{c}{period (s)} & \multicolumn{1}{c}{damping (s)}  \\
\noalign{\smallskip}
\hline
\noalign{\smallskip}
1 & $1.25\times \, 10^{-13}$ & $1.28\times\, 10^4$ & $1.37\times\, 10^4$	& $2\,874 $ & $38\,500$ &  $1\,135$ & $1\,564$  \\
2 & $2.5\times \, 10^{-13}$ &	$1.29\times\, 10^4$ & $1.47\times\, 10^4$	& $2\,961$ & $26\,600$ &  $1\,118$ & $1\,100$  \\
3 & $5\times \, 10^{-13}$ & $1.31\times\, 10^4$	& $1.68\times\, 10^4$ &  $3\,134$ & $16\,100$ &  $1\,087$ & $860$  \\
4 & $1\times \, 10^{-12}$ & $1.35\times\, 10^4$	& $2.13\times\, 10^4$ &  $3\,461$ & $9\,400$ &  $1\,053$ & $820$  \\
5 & $2\times \, 10^{-12}$ & $1.42\times\, 10^4$	& $3.07\times\, 10^4$ & $4\,160$ & $5\,600$ & 	 $1\,080 $ & $990$  \\
6 & $4\times \, 10^{-12}$ & $1.56\times\, 10^4$	& $5.04\times\, 10^4$ & $5\,600$ & $3\,800$ & 	 $1\,350$ & $1\,020 $  \\
7 & $8\times \, 10^{-12}$ & $1.84\times\, 10^4$	& $9.20\times\, 10^4$ & $7\,740 $ & $8\,100$ &  ?? & ??  \\
8 & $1.6\times \, 10^{-11}$ &	$2.41\times\, 10^4$ & $17.95\times\, 10^4$ & $10\,300 $ & $22\,200 $ &  ?? &  ??  \\
\noalign{\smallskip}
\hline
\end{tabular}
\end{center}
\end{table*}
For the coronal arcade, we choose the following parameter values: $H_{\rm d}=4\times\, 10^8$ m,
$M_{\rm d}=10^{20}$ Am and plasma-density $\rho_{\rm cor}=10^{-12}$ kg/m$^{3}$. The two flux tubes
are given by: $r_0=1.5\times\, 10^7$ m, $\Delta r_0=10^7$ m and $j_0=2\times\,
10^{-4}$ A/m$^2$. Hence the longitudinal mass density of the flux tube
proper is $\sigma_0=1.28\times\, 10^4$ kg/m, and the density of the prominence is 
$\sigma=1.35\times\, 10^4$ kg/m.
The total current is $I_0=6.5\times \, 10^{10}$ A.
Such a prominence is expected to oscillate with a horizontal period of 2777 s and a vertical
period of 1118 s. The sound speed was chosen $c_{\rm s}=128.5$ km/s, typical of the corona at
a temperature $T=10^6$ K.

We now let the initial state as defined in the previous section 
relax to a numerical equilibrium. The artificial damping term ($\partial_t \rho
\vec{v}=-\alpha \rho \vec{v}$) was
used to prevent numerical instabilities. 
We choose $\alpha=0.1$ for the first $10^4$ seconds of the
relaxation, and $\alpha=0.01$ during the latter $10^4$ seconds. Figure 
\ref{fi:relax} shows
the relaxation of the volume averaged horizontal and vertical momentum.
At the end of the
relaxation the flows in the larger part of the arcade are typically
less than 5 m/s. Along the boundary of the prominence body a rather irregular flow
field exists with velocities of 175 m/s at most.
The prominence still moves
downwards at a systematic speed of some 1 m/s (see Fig. \ref{fi:relaxy0}).
Considering both the magnitude of
the velocity perturbation applied later and the total simulation time
scale, we consider these residual flows to be unimportant. The agreement between simulations 
starting from different relaxations (with a duration of either $9\,000$ or
$20\,000$ s) confirm this.
\begin{figure*}[htb]
\centerline{\psfig{figure=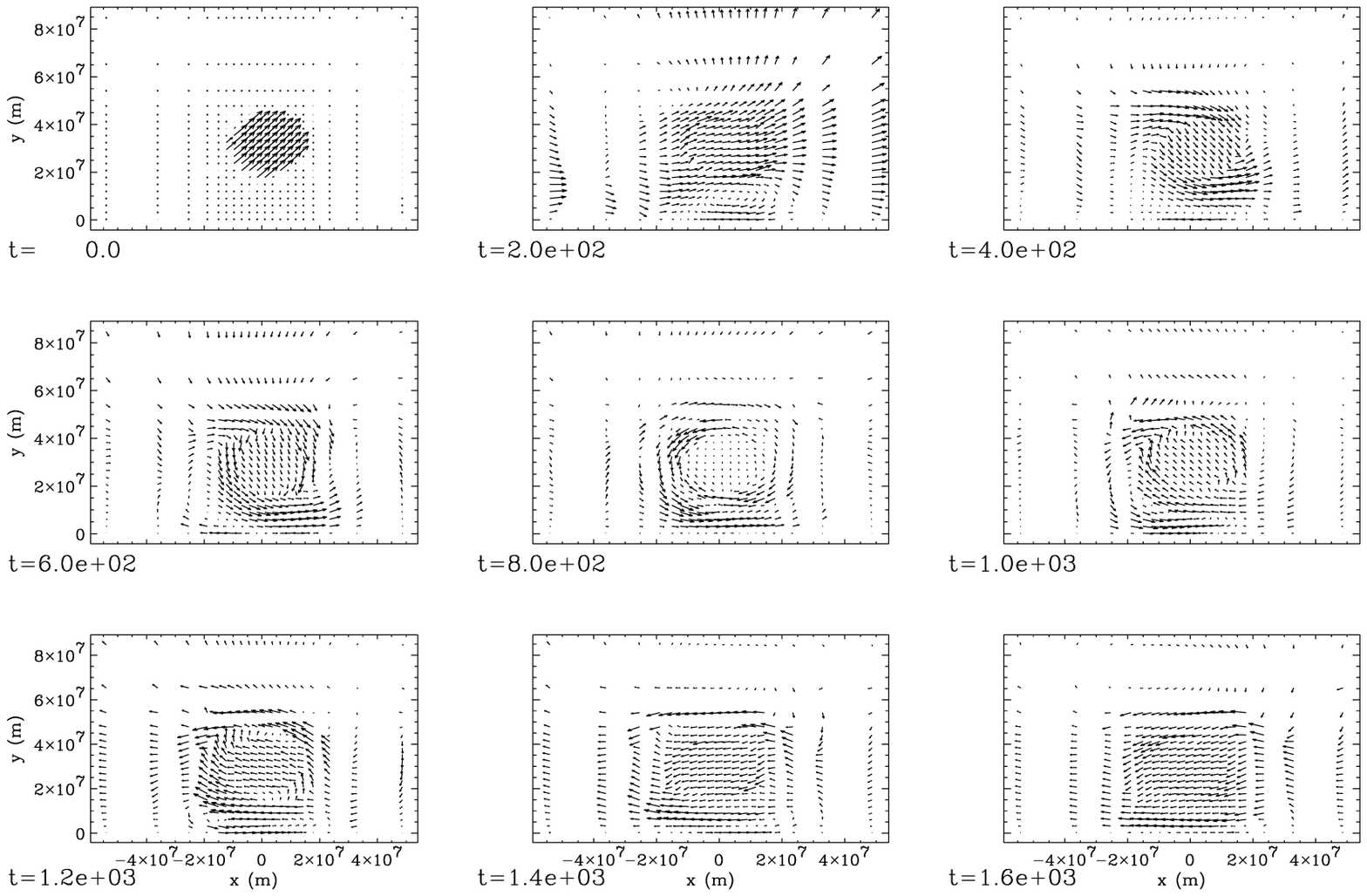,width=16cm}}
\center{\caption{\protect\footnotesize Flow field of the plasma in the prominence and ambient
corona, during the first 1600 s. Shown are the numerical results for $\rho_{\rm
cor}=10^{-12}$ kg/m$^3$, see Table\ \ref{ta:2dresults}. Maximal velocity
is 11.7 km/s.
\label{fi:FFstart}}}
\end{figure*}

The numerical equilibrium does not deviate much from the initial state. The
resulting field topology around the prominence is shown in Fig.\ \ref{fi:topol}.
The prominence itself is a region of high plasma-$\beta$, surrounded by a
region of magnetically dominated plasma (see Fig.\ \ref{fi:betava}). At large distances 
from the prominence the
plasma-$\beta$ becomes larger than unity again, as the magnetic field strength
decreases while the
coronal density is constant (due to the absence of gravity).  
The \va  
is also shown in Fig.\ \ref{fi:betava}.  It increases as 
one approaches the prominence from the corona, once inside it falls off rapidly.
The sound speed, of course, is the same everywhere (it is a free parameter of Eq.\ (\ref{eq:isomhd})).

The obtained numerical equilibrium is used to derive a series of numerical
equilibria by adding (or subtracting) a constant value from the density in each
cell. Since this does not create any additional forces (the force due to
gas pressure is $-c_{\rm s}^2 \nabla
 \rho$) a new equilibrium is found. In the resulting equilibria we disturb the inner
part of the prominence (all plasma within $1.25\times \, 10^7$ m from the
center of mass) with a velocity perturbation of 10 km/s, at an angle of
$45^o$ to the photosphere. The evolution of the system is studied for 
$10^4$ s, in some cases even for $2\times \, 10^4$ s. 


In Fig.\ \ref{fi:allm} the evolution of the volume averaged 
momentum  is shown for all eight cases considered. Initially momentum is concentrated in the
prominence, but it is redistributed throughout the corona in the subsequent
evolution.
A global oscillation is apparent,
whose properties depend strongly on coronal density.

\begin{figure*}[htb]
\centerline{\psfig{figure=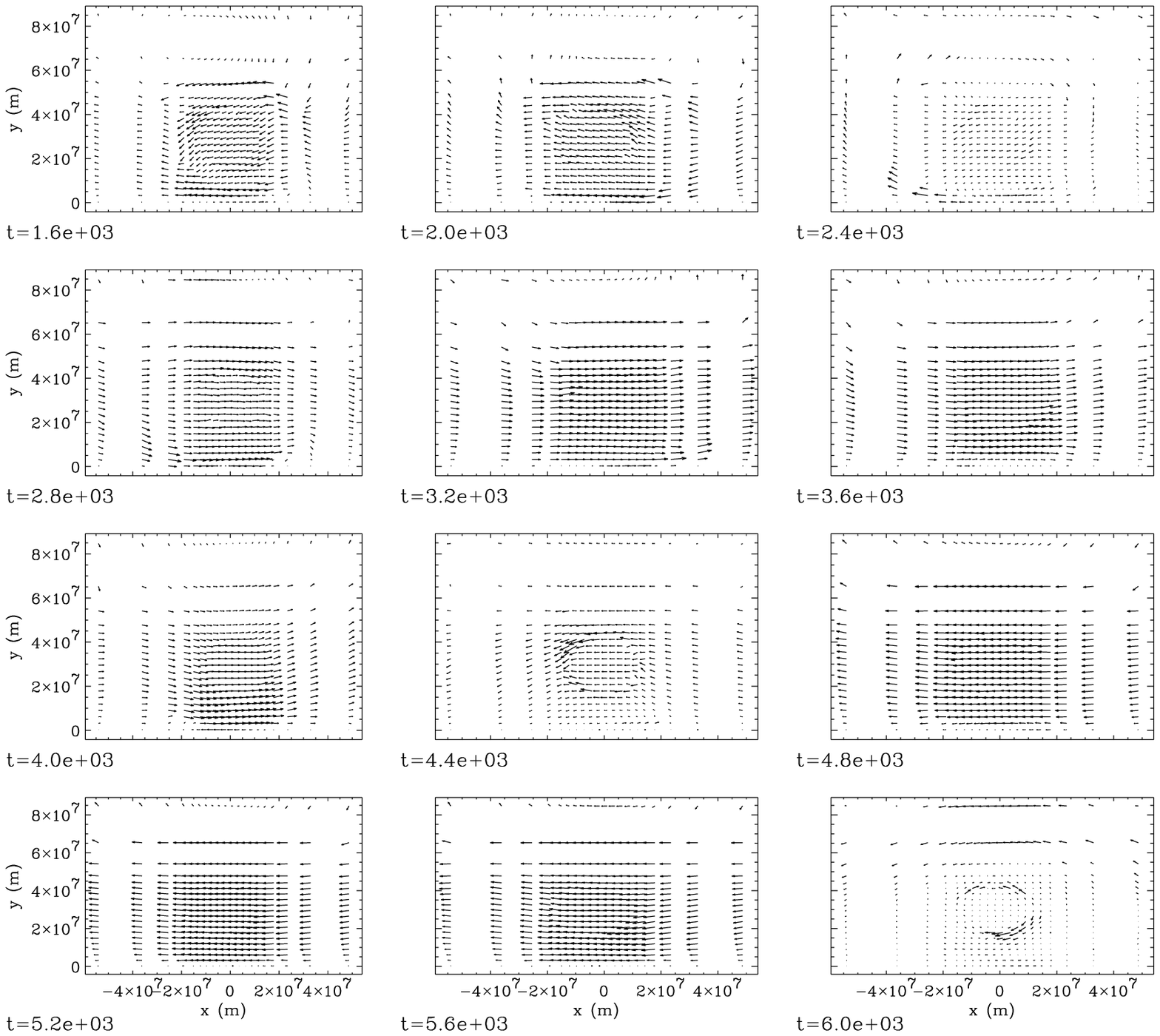,width=16cm}}
\center{\caption{\protect\footnotesize Flow field of the plasma in the prominence and ambient
corona, during the latter part of the evolution. Shown are the numerical results for $\rho_{\rm
cor}=10^{-12}$ kg/m$^3$, see Table\ \ref{ta:2dresults}. Maximal velocity
is 8.5 km/s.
\label{fi:FFend}}}
\end{figure*}

For one particular case, the flow field is shown in Figs.\ \ref{fi:FFstart} and\
\ref{fi:FFend}. Although only the velocity is shown, the
prominence stands out clearly in most graphs. It seems to move through the coronal plasma as a rigid body.
The largest velocities are usually found {\em outside} the prominence. The coronal plasma
`washes around the filament'.

In particular, we studied the motion of the prominence. This was done by
computing every so many time steps the center of longitudinal mass and current density. To
rule out the contribution of the coronal part of the grid, only densities above
a certain threshold where used. As the contrast between typical coronal and prominence
{\em current\/} densities is larger than the contrast between typical coronal and prominence
{\em mass\/} densities, the first provide a better estimate of the location of the prominence.
Nevertheless, the results are always similar. 

In essence, the motion of the
filament can be described by two decoupled damped harmonic oscillators. The
horizontal resp.\
vertical displacement of the
center of longitudinal mass or current density of the prominence may be fitted to $A {\rm
e}^{-t/T_{\rm damp}} \sin 2\pi t/P$.
The resulting periods and
damping times for the horizontal and vertical oscillations are listed in Table\
\ref{ta:2dresults}. Sometimes strong transient effects at the start of the simulation and  noise
at later times (when the velocities are smaller) cause deviations from a simple
damped harmonic oscillator. The horizontal oscillations yield, in general, better
fits. Typically the low density simulations lead to better fits than the high
density simulations. For horizontal motions we used the cases 1--6, for vertical motions we used
the cases 1--5 (see Table~\ref{ta:2dresults}). No reliable vertical time scales could be
obtained for the last two cases due to strong transients and noise.

\begin{figure}[htb]
\centerline{\psfig{figure=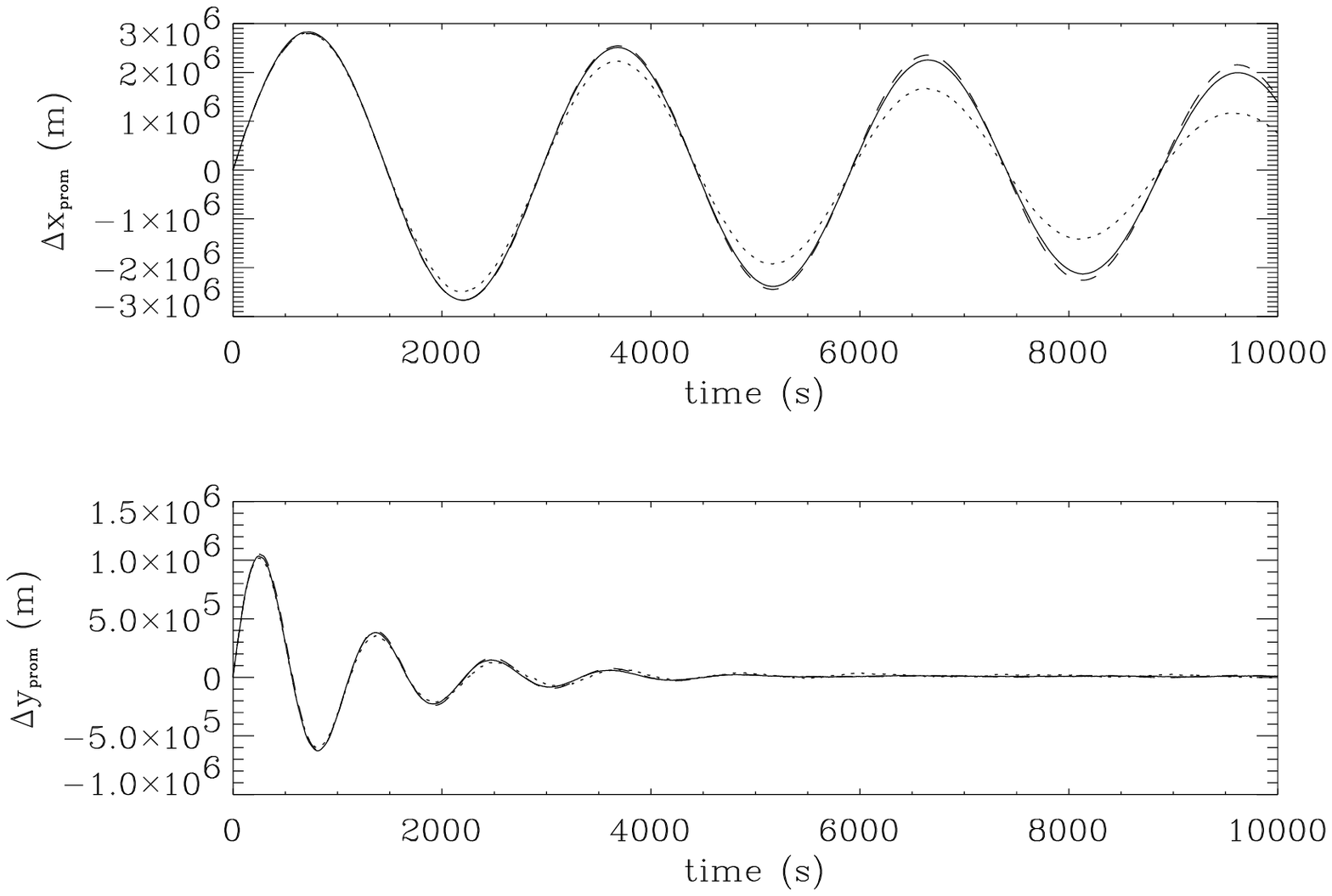,width=8.6cm}}
\center{\caption{\protect\footnotesize The center of the current density above
the threshold value. 
Dashed line: 20 km/s perturbation; solid line: 10 km/s perturbation; dotted
line: 3 km/s perturbation.
The first curve was multiplied by $0.5$, the last curve was multiplied by $3.333$ to facilitate comparison.
The top graph shows the horizontal displacement, the bottom graph the vertical
displacement. The coronal density equals
$\rho_{\rm cor}=2.5\times\, 10^{-13}$ kg/m$^3$.
\label{fi:linB1a1}}}
\end{figure}

Comparing the results with simulations where a purely vertical or horizontal perturbation was applied
shows
that the horizontal and vertical motions of the filament are actually decoupled.
This is understandable as the filament is located on the symmetry line of the
coronal field (see also van den Oord et al.\  1998). At the same time this suggests that 
non-linear effects are not important. Further evidence for the linearity
of the numerical results is found in the good fits of the
prominence oscillation curves to a damped harmonic oscillator. 
However,  simulations with 
smaller or larger perturbations (3 km/s or 20 km/s instead of 10 km/s) yield different damping times for
the horizontal motions (see Fig.\ \ref{fi:linB1a1}). For all three simulations the
horizontal  periods differ by only $0.1\%$, while the horizontal damping times differ by
$30-60\%$!
Apparently, the horizontal damping mechanism
depends non-linearly on the flow speeds. We have no explanation for this result,
but believe it to be genuine. Extensive tests with higher spatial and temporal
resolution argue strongly against the possibility of a numerical effect.
It is a strong indication that the damping
mechanism for horizontal and vertical motion differ, at least in their dependencies on the flow
field.
  
\begin{figure}[htb]
\centerline{\psfig{figure=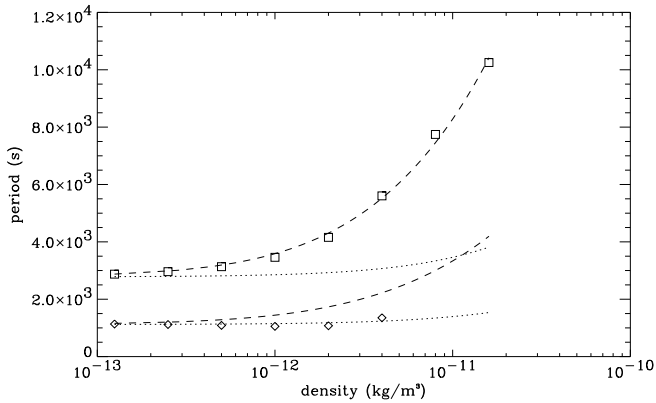,width=8.6cm}}
\center{\caption{\protect\footnotesize Periods for the horizontal (squares) and vertical
(diamonds)
oscillations. The dotted lines are the quasi-stationary periods for a prominence loaded with mass
$\sigma=\sigma_0+\pi r_0^2 \rho_{\rm cor}$. The top dashed line is the model for horizontal
oscillations assuming that the actual oscillating body is larger than just the prominence proper.
The bottom dashed line is the model for vertical oscillations using the
effective longitudinal mass density $\sigma_{\rm eff}$ as determined from the
horizontal oscillations. 
\label{fi:period}}}
\end{figure}

\begin{figure}[htb]
\centerline{\psfig{figure=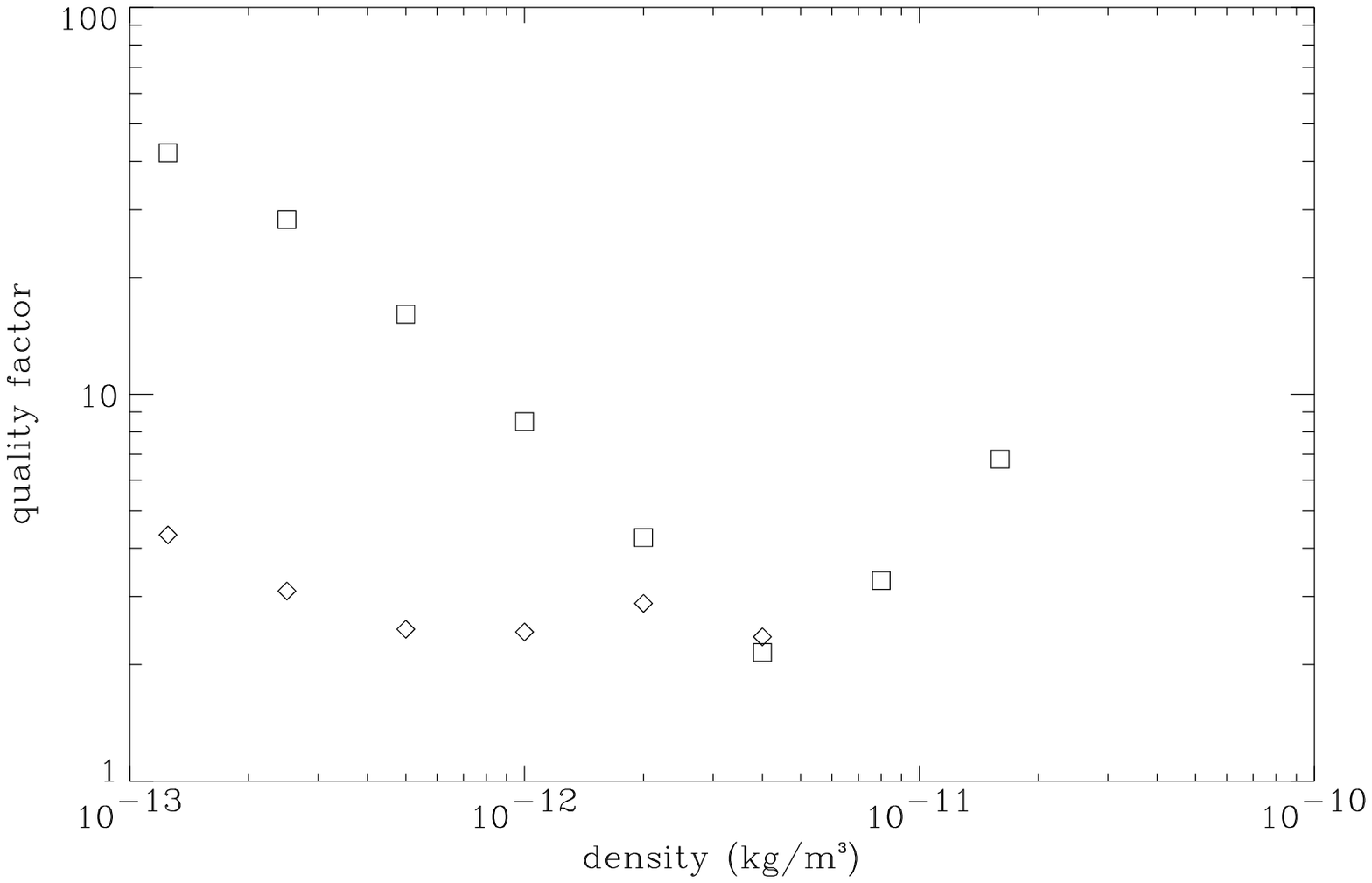,width=8.6cm}}
\center{\caption{\protect\footnotesize Quality factors ($Q=\pi T_{\rm damp}/P$) for the horizontal
(squares) and vertical (diamonds)
oscillations. 
\label{fi:quality}}}
\end{figure}
The horizontal periods show a strong dependence on the coronal
density, while the vertical periods are almost constant (Fig.\ \ref{fi:period}).
The quality factors (Fig. \ref{fi:quality}) of the horizontal
oscillation are typically larger than four, and hence damping contributes little
to the
oscillation frequency (see Eq.\ (\ref{eq:horfreq})). 
The dotted lines (in Fig.\ \ref{fi:period})
represent the undamped quasi-stationary period for a prominence with a longitudinal
mass density of $\sigma=\sigma_0+\pi r_0^2 \rho_{\rm cor}$. Clearly this is a bad
fit to the numerical results for horizontal oscillations. Let us assume that, due to the magnetic structure, the actual oscillating body
has a radius $r_{\rm eff}>r_0$. The actual oscillating body now
has an effective longitudinal mass density $\sigma_{\rm eff}=\sigma_0+ \pi r_{\rm eff}^2 \rho_{\rm cor}$.
From Eq.\ (\ref{eq:horfreq}) we obtain
\begin{equation}
\frac{1}{\rho_{\rm cor}} \left( \frac{2 \mu_0 M_{\rm d}^2}{\pi^3}
\frac{y_0}{(y_0+H_{\rm d})^5} P_x^2 - \sigma_0 \right) = 
\pi r_{\rm eff}^2 
\end{equation}
\noindent
the left-hand-side of which can be fitted to a function of the form $C \rho_{\rm
cor}^{\gamma}$.
We find that $r_{\rm eff}=5.0\times\,
10^7 (\rho_{\rm cor}/10^{-12})^{0.04}$ m, weakly dependent on the coronal density. The typical
radius $r_{\rm eff}=5.0\times\, 10^7$ m agrees well with the location of the magnetic surface across which
the connectivity of the field lines changes from closed field lines in the corona to field lines
anchored in the photosphere. An even better approximation to this surface is obtained by taking into account
the effect of induced mass as described by Landau \& Lifschitz (1989, p. 29; see also Lamb 1945).
For an incompressible, potential
hydrodynamical flow  the effective mass is the sum of the mass of the actual
oscillating body ($\sigma_0+\pi r_0^2 \rho_{\rm cor}$) plus 
the mass of the fluid displaced by the body. In that case 
$\sigma_{\rm eff}=\sigma_0+ 2 \pi r_{\rm eff}^2 \rho_{\rm cor}$ (note the factor 2!) and we find 
$r_{\rm eff}=3.5\times\, 10^7 (\rho_{\rm cor}/10^{-12})^{0.04}$ m.

\begin{figure}[htb]
\centerline{\psfig{figure=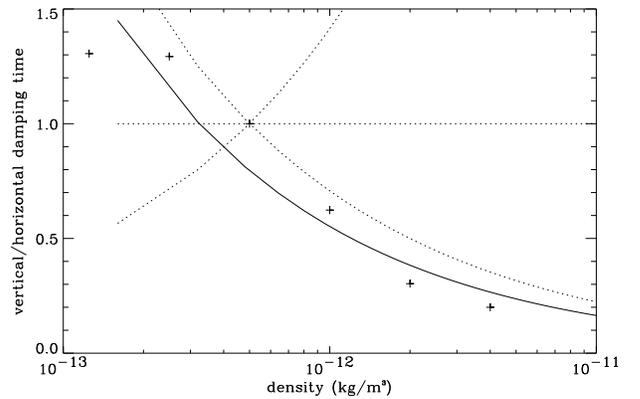,width=8.6cm}}
\center{\caption{\protect\footnotesize Ratio of the damping times of vertical and
horizontal
oscillations. The solid line ($\propto \rho_{\rm
cor}^{-0.53}$) is a fit to the data points (plus signs).  The dotted lines are fits to $c_x/c_y \propto
\rho_{\rm cor}^{-0.5}, 1$ or $\rho_{\rm cor}^{0.5}$ using the data point for $\rho_{\rm cor}=5\times\,
10^{-13}$ kg/m$^3$ as a gauge. 
\label{fi:damping}}}
\end{figure}

In Fig.\ \ref{fi:damping}, we have
plotted the ratio of vertical damping time to horizontal damping time. From Eqs.\
 (\ref{eq:damping}), (\ref{eq:hordamp}) and\ (\ref{eq:vertdamp}) this ratio 
is proportional to $c_x/c_y$.
Here $c_x$ is the typical wave speed for perturbations travelling parallel to the photosphere and $c_y$ is 
the typical wave speed for perturbations travelling perpendicular to the photosphere.
For simplicity, we assume that the cross sections 
for both directions have the same $\rho_{\rm cor}$ dependence, but given the azimuthal invariance of the
prominence flux tube this is probably a  fair approximation. 
Now $c_x$ and $c_y$ are equal to either the slow cusp speed $c_{\rm T}$, the
Alfv\'en speed $c_{\rm A}$ or
the fast speed $c_{\rm f}$. In a low plasma-$\beta$ environment, the cusp speed equals
the sound speed $c_{\rm s}$, while the fast speed equals the Alfv\'en speed
$c_{\rm A}$.
Hence $c_x/c_y \propto
\rho_{\rm cor}^{-0.5}, 1$ or $\rho_{\rm cor}^{0.5}$ depending on whether slow,
Alfv\'en or fast
wave emission prevails in a certain direction. From the data, we find $c_x/c_y \propto \rho_{\rm
cor}^{-0.53}$ which strongly suggests that slow wave emission damps horizontal
motions, while Alfv\'en or fast
wave emission damps vertical motions. Since the prominence moves vertically to
the field lines in the latter case, fast waves are more likely.

\begin{figure}[htb]
\centerline{\psfig{figure=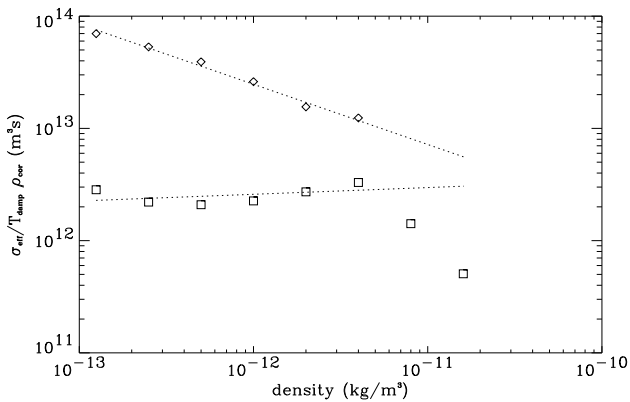,width=8.6cm}}
\center{\caption{\protect\footnotesize The effective longitudinal mass density $\sigma_{\rm eff}$ over 
the damping time $T_{\rm d}$ and coronal density $\rho_{\rm cor}$ for horizontal (squares) and vertical (diamonds) oscillations. The dotted lines are 
model fits to Eq.~(\ref{eq:dampmodel}). 
\label{fi:wave}}}
\end{figure}

The conclusions regarding damping mechanisms can be substantiated further by comparing the
relevant time scales for horizontal and vertical oscillations independently to the Eqs.\
(\ref{eq:hordamp}) and\ (\ref{eq:vertdamp}) (see Fig.\ \ref{fi:wave}). 
For the mass, we use the effective longitudinal mass density $\sigma_{\rm eff}$ as obtained
previously. 
From Eq.\ (\ref{eq:damping}) and either Eq.\ (\ref{eq:hordamp}) or\
(\ref{eq:vertdamp}) we find
\begin{equation}
\label{eq:dampmodel}
\frac{\sigma_{\rm eff}}{\rho_{\rm cor} T_{\rm damp}} = A c, 
\end{equation}
\noindent
the left-hand-side of which can be fitted to $C \rho_{\rm cor}^{\gamma}$.
Here $A$ is the cross-section perpendicular to the direction of motion.
For horizontal oscillations we find $c_x \propto \rho_{\rm cor}^{0.06 \pm 0.06}$, and $A \approx 2\times\, 10^7$ m$^2$. 
This suggests excitation of slow waves by horizontal motion of the prominence.
For vertical oscillations $c_y \propto \rho_{\rm cor}^{-0.54 \pm 0.04}$ which
suggests excitation of fast waves by vertical motion of the prominence.
Since the \va varies in space, it
is not possible to derive a value of $A$ from the product of $Ac$.

The cross-section $A$ for the horizontal damping mechanism yields a dimension
for the oscillating body smaller than the extended size $r_{\rm eff}$ as obtained from the
horizontal periods. 
This may be due
to field line curvature. Since the size of the oscillating `solid' body is determined by the field
topology, Lorentz forces cause the waves that carry away momentum. However, Lorentz forces only act
perpendicular to, not parallel to the field lines. Thus only perturbations of
magnetic field with a strong vertical component can excite horizontally
travelling waves. 

\section{Summary and conclusions}
\label{se:summary}

We have made a numerical investigation of prominence oscillations, by solving the isothermal MHD equations
in two dimensions. First we computed a prominence
equilibrium that is very similar to the Kuperus-Raadu (inverse polarity) topology. 
However, in our numerical equilibrium the prominence is not infinitely thin, but instead well resolved.
From this equilibrium, we derived other equilibria with different coronal plasma densities.
We then perturbed the system by instantaneously adding momentum to the prominence mass and followed the ensuing
oscillations. The dependence of the characteristic time scales (periods and damping
times) on the coronal plasma density was analyzed in terms of a solid body moving
through a fluid. To our knowledge this is the first attempt at numerically simulating prominence
oscillations.

In our numerical model, we ignored the effects of gravity and thermodynamics, for the sake of
clarity and practicality. Also, up to now the numerical boundary conditions we are using  do not
seem to allow a stable gravitationally stratified corona.

We believe, however, that
gravity and thermodynamics do not contribute significantly to the physics of the system. Gravity is of  small consequence 
for the equilibrium of a Kuperus-Raadu prominence as detailed by van Tend \& Kuperus (1978). Force
balance is due to two Lorentz forces: one due to the coronal magnetic arcade, the other due to the
photospheric flux conservation. The gravitational force can be ignored when describing global equilibrium.
Furthermore, the scale height of a corona of
$T=10^6$ K is $3\times\, 10^8$ m, which is larger than the typical vertical size of 
prominences. The inclusion of gravity would change the appearance of the prominence into a slab,
with prominennce matter accumulating in the pre-existing (!) dips in the magnetic field lines. The
magnetic field configuration would hardly change.

Likewise, the absence of a thermally structured corona and prominence seems of only minor influence.
The most important effect would be that a cool prominence will be heavier than a prominence of coronal
temperatures. In our model this is balanced by the absence of a longitudinal field. 
As a consequence, the pinching effect of the azimuthal prominence field has to be balanced 
by
gas pressure only (in real prominences the longitudinal field pressure contributes significantly). The
total longitudinal mass density of our prominence ($\sim 1.3\times 10^4$ kg/m$^3$) agrees well with
that of a slab of height $5\times\, 10^7$ m, width $6\times\,10^6$ m and density $5\times\,10^{-11}$
kg/m$^3$. Oliver \& Ballester (1996) studied the influence of the prominence-corona transition
region (characterized by strong temperature gradients) and found that it mainly influences the
prominence internal oscillations, but not the global oscillations.

We point out that the inclusion of both gravity and thermodynamics would give the prominence proper the
appearance of a cool slab, suspended in the  dips of the field lines belonging to the prominence current.
Presumably this will not change the global oscillation discussed in this paper, since that mode is
determined by the overall field structure. 

The results indicate that, for typical coronal densities ($\rho_{\rm
cor}=10^{-13}-10^{-12}$ kg/m$^3$), the prominence structure can indeed be viewed
as a solid body moving through a fluid. However, the mass of this solid body is determined by the
magnetic topology, not the prominence proper. In particular, the mass of the body seems to be determined by
coronal field lines that enclose the prominence proper. In a low plasma-$\beta$
environment this is
to be expected. As a consequence, the total mass of the oscillating solid body is
larger than the mass of the prominence proper. 

Due to the symmetry of the system, horizontal and vertical prominence oscillations
decouple. These oscillations can each be interpreted as the motion of a damped harmonic
oscillator. The horizontal periods and the horizontal and vertical damping
times can be explained by assuming that the actual oscillating structure is
larger than the prominence proper, due to the magnetic field topology. Only the
vertical periods do not agree with this model. They are nearly constant for
different values of the coronal density and are best modelled by the oscillation
of the prominence proper in a corona with vanishing plasma density ($\rho_{\rm
cor} \downarrow 0$). For realistic coronal densities, the horizontal periods
change with $3-15\%$ at most when taking the actual oscillating structure into
account. However, for lower prominence longitudinal mass densities $\sigma$ in a stronger coronal background field
(and hence larger prominence currents $I_0$), the effect will be much more
pronounced.

Vertical oscillations lead to the emission of fast waves that carry momentum
away from the prominence and damp the oscillation. Horizontal oscillations, on the other
hand, lead to the emission of slow waves. These will damp the horizontal oscillation, but 
less effectively than fast waves ($Q_x>Q_y$). 

The difference in wave emission between horizontal and vertical oscillations can be understood in terms of the 
coronal arcade in which the prominence is embedded. Due to the large scale height, the arcade field 
is close to horizontal near the prominence. Waves that travel in the vertical direction (up or down)
therefor travel more or less perpendicular to the field lines and must be fast waves. Waves that
travel in the horizontal direction travel along the field lines. They could be either
Alfv\'en waves or magneto-acoustic waves. As regards excitation of the waves, it is obvious that the
vertical motion of a prominence across field lines that are nearly horizontal will compress both gas
and magnetic field  and thus set off fast waves. The excitation of the magneto-acoustic slow waves
(for our analysis strongly suggests they are slow waves) is not as well understood. But apparently
the prominence acts as a piston during horizontal motions along magnetic field lines of the arcade.

We surmise that for smaller scale height of the arcade the slow waves might be replaced by fast
waves. 

\begin{acknowledgements} N.A.J. Schutgens was financially supported by the Netherlands
 Organisation for Scientific Research (NWO) under grant nr. 781-71-047. 
 He gratefully acknowledges stimulating discussions with Max Kuperus 
 and  Bert van den Oord. The Versatile Advection Code (VAC) was developed by 
 G. T\'{o}th as part of the project on `Parallel Computational Magneto-Fluid
 Dynamics', funded by the Netherlands
 Organisation for Scientific Research (NWO) Priority Program on Massively
 Parallel Computing, while he was working at the Astronomical Institute of
 Utrecht University. G. T\'{o}th currently receives a post-doctoral fellowship
 (D~25519) from the Hungarian Science Foundation (OTKA).
\end{acknowledgements}

\end{document}